\documentclass[aps,pra,reprint,groupedaddress,showpacs]{revtex4-1}

\usepackage{amsfonts,amssymb,amsmath}
\usepackage{amstext}
\usepackage{graphicx}
\usepackage[caption=false]{subfig}
\usepackage{textcomp}

\newcommand{\LandLifsGilb}{Landau--Lifshitz--Gilbert}

\newcommand{\Bloch}{Bloch}
\newcommand{\Doring}{D\"{o}ring}
\newcommand{\Fourier}{Fourier}
\newcommand{\Neel}{N\'{e}el}
\newcommand{\Peierls}{Peierls}
\newcommand{\Poisson}{Poisson}
\newcommand{\Novoselov}{Novoselov}

\newcommand{\Kosevich}{Kosevich}

\DeclareMathOperator{\sech}{sech}
\newcommand{\Or}{\mathcal{O}}
\newcommand{\e}{e}

\newcommand{\figref}[1]{Fig.~\ref{#1}}
\newcommand{\Figref}[1]{Figure~\ref{#1}}
\newcommand{\equaref}[1]{Eq.~\eqref{#1}}
\newcommand{\Equaref}[1]{Equation~\eqref{#1}}

\newcommand{\explcite}[1]{Ref.~[\onlinecite{#1}]}
\newcommand{\Explcite}[1]{Reference~[\onlinecite{#1}]}

\begin{document}

\title{Motion of domain walls and the dynamics of kinks in the magnetic \Peierls{} potential}

\author{F. J. Buijnsters}
\email[]{F.Buijnsters@science.ru.nl}
\author{A. Fasolino}
\author{M. I. Katsnelson}
\affiliation{Institute for Molecules and Materials, Radboud University Nijmegen, Heyendaalseweg~135, 6525~AJ Nijmegen, Netherlands}

\date{July 25, 2014}

\begin{abstract}
We study the dynamics of magnetic domain walls in the \Peierls{} potential due to the discreteness of the crystal lattice.
The propagation of a narrow domain wall (comparable to the lattice parameter) under the effect of a magnetic field proceeds through the formation of kinks in its profile.
We predict that, despite the discreteness of the system, such kinks can behave like sine-Gordon solitons in thin films of materials such as yttrium iron garnets, and we derive general conditions for other materials.
In our simulations we also observe long-lived breathers.
We provide analytical expressions for the effective mass and limiting velocity of the kink in excellent agreement with our numerical results. 
\end{abstract}

\maketitle

\paragraph*{Introduction.}
The statics and dynamics of magnetic domain walls have been studied intensively because they determine the most important technical characteristics of magnetic materials, such as magnetization curves and hysteresis \cite{Aharoni2001}. Recently, there has been a revival of interest in this field due to the development of new techniques to manipulate magnetization, such as current-induced spin transfer torque \cite{Slonczewski1996,Berger1984} and optical control \cite{Kirilyuk2010}. These developments open new perspectives for magnetic data storage \cite{Parkin2008} and call for a deeper understanding of the elementary processes associated with domain-wall motion.
Traditionally, magnetization profiles are described using continuum models (micromagnetics) \cite{Brown1963}.
However, it is increasingly being recognized that the \emph{discrete nature} of the crystal lattice can play an important role in both the statics and the dynamics of magnetic topological defects including domain walls \cite{Novoselov2003}, (nano)skyrmions \cite{Heinze2011}, and \Bloch{} points \cite{Andreas2014}.

\begin{figure}
  \includegraphics[width=\columnwidth]{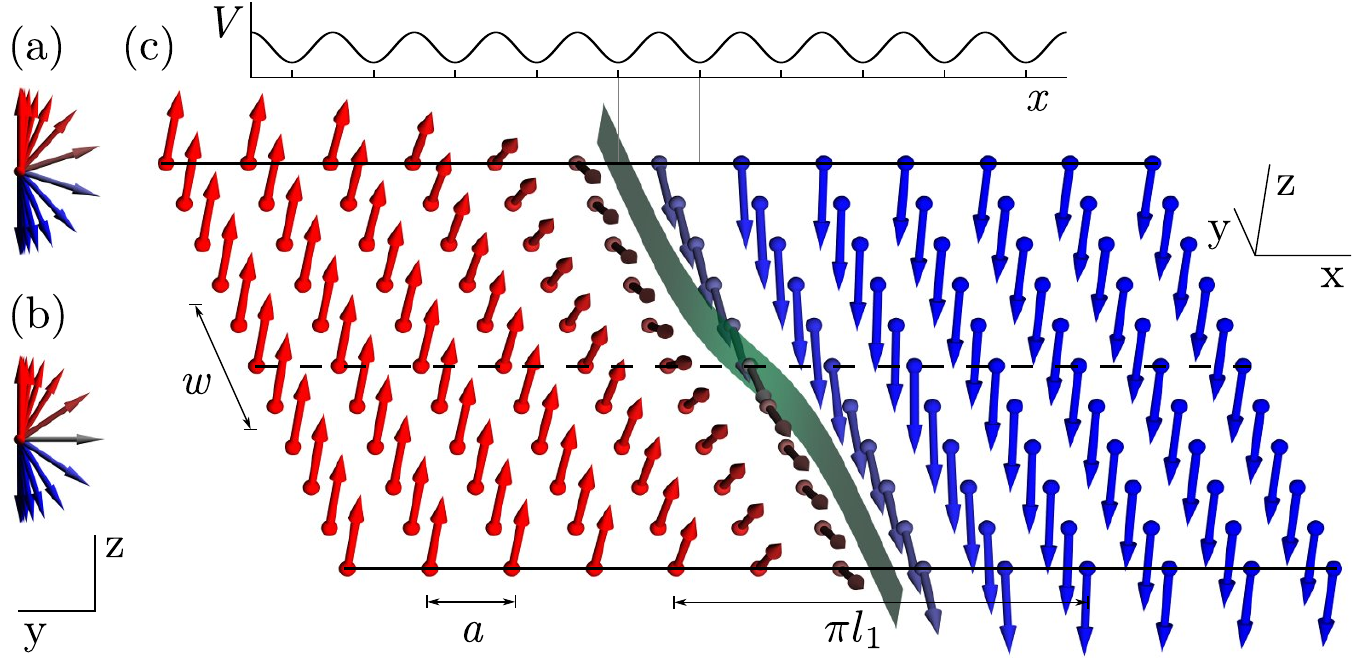}
  \caption{\label{fig:kink}(color online). A Bloch domain wall in a thin film with perpendicular anisotropy.
  (a,b) Side views of the domain wall in an energetically favorable [(a), solid lines in (c)] and unfavorable [(b), dashed line in (c)] position in the \Peierls{} potential (top). (c) Domain wall with an antikink.
  The center of the domain wall is indicated as a translucent strip.
  }
\end{figure}

It has been predicted \cite{vandenBroek1971,Hilzinger1972,Barbara1994} that if the characteristic thickness of a domain wall is comparable to the lattice parameter, it may become trapped in a favorable position between two crystallographic planes, as shown in \figref{fig:kink}.
The energy of the domain wall as a function of the position $x$ of its center shows a pattern of peaks and valleys with a periodicity $a$ determined by the lattice. The analogous effect for dislocations is known in the field of crystal plasticity as the \Peierls{} potential or \Peierls{} relief \cite{Hirth1982}.
\Novoselov{} {\it et al.}\ \cite{Novoselov2003} confirmed the existence of the magnetic \Peierls{} potential in thin films of yttrium iron garnet (YIG) that combine the very large unit cell (80 atoms) with relatively strong perpendicular anisotropy. Jumps of the domain wall between valleys of the \Peierls{} potential were detected.

In two- or three-dimensional systems, domain walls are not necessarily flat and different areas may lie in different \Peierls{} valleys, as shown in \figref{fig:kink}(c). The boundaries (kinks) between such areas are energetically unfavorable, but once present they can slide freely along the domain wall, effectively moving it in steps of distance $a$. Dislocations in crystals are known to move in a similar way \cite{Hirth1982, Gornostyrev1999}. Measurements of AC magnetic susceptibility provide evidence for kinks in domain walls (\emph{DW-kinks}) in YIG thin films \cite{Novoselov2003}.

If the \Peierls{} potential is significant, the motion of a domain wall is determined by the dynamics of DW-kinks.
A crucial question is what happens when two kinks of opposite sign collide: either they pass through each other, like sine-Gordon solitons \cite{Dodd1982}, or they annihilate.
In the former case, a domain wall can jump to another \Peierls{} valley and propagates more efficiently.
While many authors have discussed the dynamics of kinks in dislocations \cite{Gornostyrev1999,Kosevich2005,Swinburne2013A}, we point out that DW-kinks are different in essential respects \cite{supmat}.
In this Letter, we show that DW-kinks can display solitonic behavior.
We derive necessary conditions in terms of the characteristic lengthscales of the system, and
we predict the existence of long-lived breathers (bound kink--antikink pairs \cite{Dodd1982}) in thin films of materials such as YIG.
We also find that DW-kinks possess inertia, somewhat analogous to the \Doring{} effective mass \cite{Doring1948}, and we derive an expression for the DW-kink mass valid for both solitonic and nonsolitonic cases,
in excellent agreement with numerical simulations.

\paragraph*{Model.}

We consider a magnetic thin film of thickness $L$ with perpendicular anisotropy, assuming for simplicity a simple cubic lattice.
We argue that for our purposes the film may be considered as effectively two-dimensional if $L \ll w$, where $w$ is the characteristic width of the DW-kink (determined below). While in the direction normal to the domain wall the magnetization profile varies on the much shorter scale of the exchange length, in this direction the domain wall cannot move freely and remains in a valley of the \Peierls{} potential.

We model the dynamics of the localized magnetic moments, described by unit vectors  $\mathbf{m}_{ij}$, using the \LandLifsGilb{} (LLG) equation \cite{Aharoni2001},
\begin{equation}\label{eq:llg}
\frac{d\mathbf{m}_{ij}}{dt} =
\frac{\lvert \gamma \rvert}{a^2 M_\text{S}} \mathbf{m}_{ij} \times
 \nabla_{\mathbf{m}_{ij}}\mathcal H
 + \alpha \mathbf{m}_{ij} \times \frac{d\mathbf{m}_{ij}}{dt}\text{,}
\end{equation}
where $\gamma$ is the gyromagnetic ratio, $M_\text{S}$ is the saturation magnetization and $\alpha$ is the dimensionless Gilbert damping parameter. The lattice sites are located at $(x,y)=((i+\tfrac{1}{2})a,(j+\tfrac{1}{2})a)$, $i,j\in\mathbb{Z}$, where $a$ is the lattice parameter.
The Hamiltonian $\mathcal H$ is given by
\begin{multline}\label{eq:hamiltonian}
\mathcal H = \sum_{ij} a^2 \bigl( - \frac{2A}{a^2}[\mathbf{m}_{ij} \cdot \mathbf{m}_{(i+1)j} + \mathbf{m}_{ij} \cdot \mathbf{m}_{i(j+1)}] \\
 - K_1 {(\mathbf{m}_{ij} \cdot \hat{\mathbf{z}})}^2 + K_2 {(\mathbf{m}_{ij} \cdot \hat{\mathbf{x}})}^2  - M_\text{S} \mathbf{H}_\text{app} \cdot \mathbf{m}_{ij} \bigr) \text{.}
\end{multline}
Here $A$ represents the exchange parameter, $K_1>0$ the anisotropy for the easy axis $\hat{\mathbf{z}}$, $K_2$ the in-plane anisotropy, and $\mathbf{H}_\text{app}$ the applied field.
The corresponding continuum model gives an exchange length $l_1 = \sqrt{A/K_1}$, a Bloch domain-wall width of $\pi l_1$, and a Bloch domain-wall energy of $\epsilon_1 = 4\sqrt{A K_1}$ per unit area.
We consider a domain wall oriented normal to the $x$ direction, as shown in \figref{fig:kink}(c).
The effect of dipolar interactions is taken into account through the second anisotropy parameter $K_2 = 2\pi M_\text{S}^2$, which penalizes magnetization in the $x$ direction. 
This approximation, exact for planar magnetization profiles $\mathbf{m}(x)$ \cite{Aharoni2001,Dodd1982}, has been used in other contexts where the domain wall is only approximately flat \cite{Thiele1973B}. It can be justified here because we assume $l_1 \ll L$ and $l_1 \ll w$.

\paragraph*{Statics.}

Minimizing the atomistic Hamiltonian \eqref{eq:hamiltonian} under the constraint of a fixed domain-wall center $x$ gives a \Peierls{} potential of the form \cite{vandenBroek1971, Hilzinger1972,Egami1973A}
\begin{equation}
V(x) = V_0  ( 1 - \cos 2\pi x/a )\text{.}
\end{equation}
This sinusoidal shape is known to be insensitive to the crystal structure up to exponentially small corrections \cite{Hilzinger1972,Kosevich2005}.
The strength $V_0$ depends very sensitively on the ratio between the domain-wall width $\pi l_1$ and the distance $a$ between equivalent crystallographic planes; we have $V_0 \sim e^{-\pi^2 l_1 / a}$ \cite{Hilzinger1972}.
Noticeable effects require $l_1 \lesssim 2.5a$.

Static configurations of domain walls with kinks and the thermally activated formation of kink loops were studied theoretically in \explcite{Egami1973A}.
Let us describe the profile of the domain-wall center, shown as a strip in \figref{fig:kink}(c), as a function $x(y)$, which we define via the average magnetization $m_z$ on a line of constant $y$.
The equilibrium profile of a single kink is given by \cite{Egami1973A}
\begin{equation}\label{eq:kinksolution}
x(y) = \frac{2a}{\pi} \arctan {\Bigl[ \exp  \Bigl(\pi \frac{y-y_0}{w}\Bigr)  \Bigr]}\text{,}
\end{equation}
where $y_0$ is the center of the kink and $w = \tfrac{1}{2} a \sqrt{\epsilon_1 / V_0}$ is its characteristic width, which arises from a competition between the \Peierls{} potential and the exchange energy. The kink energy per unit length is $\lambda = 4a\sqrt{\epsilon_1 V_0} / \pi$ \cite{Egami1973A}.

We express the kink width in the experimentally accessible quantity $H_\text{c}$, the coercive field of the \Peierls{} barrier:
\begin{equation}\label{eq:width}
w = \sqrt{\frac{\pi a \epsilon_1}{4 M_\text{S} H_\text{c}}}\text{.}
\end{equation}
Taking experimental parameters from \explcite{Novoselov2003}, we find
$l_1 = 3.6\,\text{nm} = 2.0a$, 
$\epsilon_1 = 2.0 \,\text{erg}\,\text{cm}^{-2}$,
$\lambda = 2.5\times10^{-10} \,\text{erg}\,\text{cm}^{-1} = 7\times10^{-4} a \epsilon_1$,
and $w = 1.6\,\text{{\textmu}m} \approx 900a$.

\paragraph*{Dynamics.}

We express the Hamiltonian \eqref{eq:hamiltonian} in terms of the collective coordinates $x(y),\vartheta(y)$ of the domain wall.
The angle $\vartheta$, canonically conjugate to $x$, represents the in-plane orientation of the magnetization near the center of the domain wall \cite{Tatara2008}. We define $\vartheta = 0,\pi$ for a \Bloch{} domain wall and $\vartheta = \pm\pi/2$ for a \Neel{} domain wall.
For $\mathbf{H}_\text{app} = 0$ and in the limit of small $\vartheta$, we get
\begin{equation}\label{eq:hamapprox}
\mathcal{H} \approx
  \int \biggl[
  \frac{\epsilon_1}{2} \frac{K_2}{K_1} \vartheta^2
  + \frac{\epsilon_1}{2} \Bigl( \frac{\partial x}{\partial y} \Bigr)^2
  + V(x)
  + \frac{\epsilon_1 l_1^2}{2} \Bigl( \frac{\partial \vartheta}{\partial y} \Bigr)^2
  \biggr] dy
\text{.}
\end{equation}
We assume $w \gg a$, so that the system is effectively continuous in $y$.
Since $l_1 \sim a$, higher-order terms of the exchange energy may give corrections to \equaref{eq:hamapprox}, but we find that such corrections are relatively small \cite{supmat}.

The \Poisson{} brackets for $x(y),\vartheta(y)$ are given by
$\lbrace x(y),\vartheta(y') \rbrace =  | \gamma | / (2 M_\text{S}) \, \delta(y-y') $ and $\lbrace x(y),x(y') \rbrace = \lbrace \vartheta(y),\vartheta(y') \rbrace = 0$.
Taking into account Gilbert damping in the small-$\vartheta$ limit, we get equations of motion \cite{Tatara2008}
\begin{subequations}\label{eq:geneom}
\begin{align}
\dot{x}(y)
  & = \frac{|\gamma|}{2M_\text{S}} \frac{\delta \mathcal{H}}{\delta \vartheta(y)}
  + \alpha l_1 \dot{\vartheta}(y)
  \text{,} \\
\dot{\vartheta}(y) 
  & = - \frac{|\gamma|}{2M_\text{S}} \frac{\delta \mathcal{H}}{\delta x(y)}
  - \frac{\alpha}{l_1} \dot{x}(y)
  \text{,}
\end{align}
\end{subequations}
where a dot denotes the time derivative.

\paragraph*{Solitonic behavior.}

Let us define a second ``exchange length'' $l_2 = \sqrt{A/K_2}$.
Neglecting the term in $\partial \vartheta / \partial y$ in \equaref{eq:hamapprox}, \equaref{eq:geneom} with $\alpha = 0$ reduces to the sine-Gordon equation,
\begin{equation}\label{eq:sG}
T^2 \frac{\partial^2 \varphi}{{\partial t}^2} - Y^2 \frac{\partial^2 \varphi}{{\partial y}^2} + \sin \varphi = 0 \text{,}
\end{equation}
where we define $\varphi=2\pi x / a$, $Y=w/\pi$, and
\begin{equation}
T=\frac{a^2}{2 \pi^2 E} \frac{M_\text{S} }{ |\gamma|} \frac{l_2}{l_1} \text{,}
\end{equation}
with $E = \lambda / 8$ (characteristic energy scale).
The sine-Gordon equation is one of the very few mathematical models that allow for truly solitonic behavior \cite{Dodd1982}. This means that DW-kinks behave like solitons only to the extent that \equaref{eq:sG} is a good approximation.
We now investigate under which conditions this is the case.

\begin{figure}
  \includegraphics[width=\columnwidth]{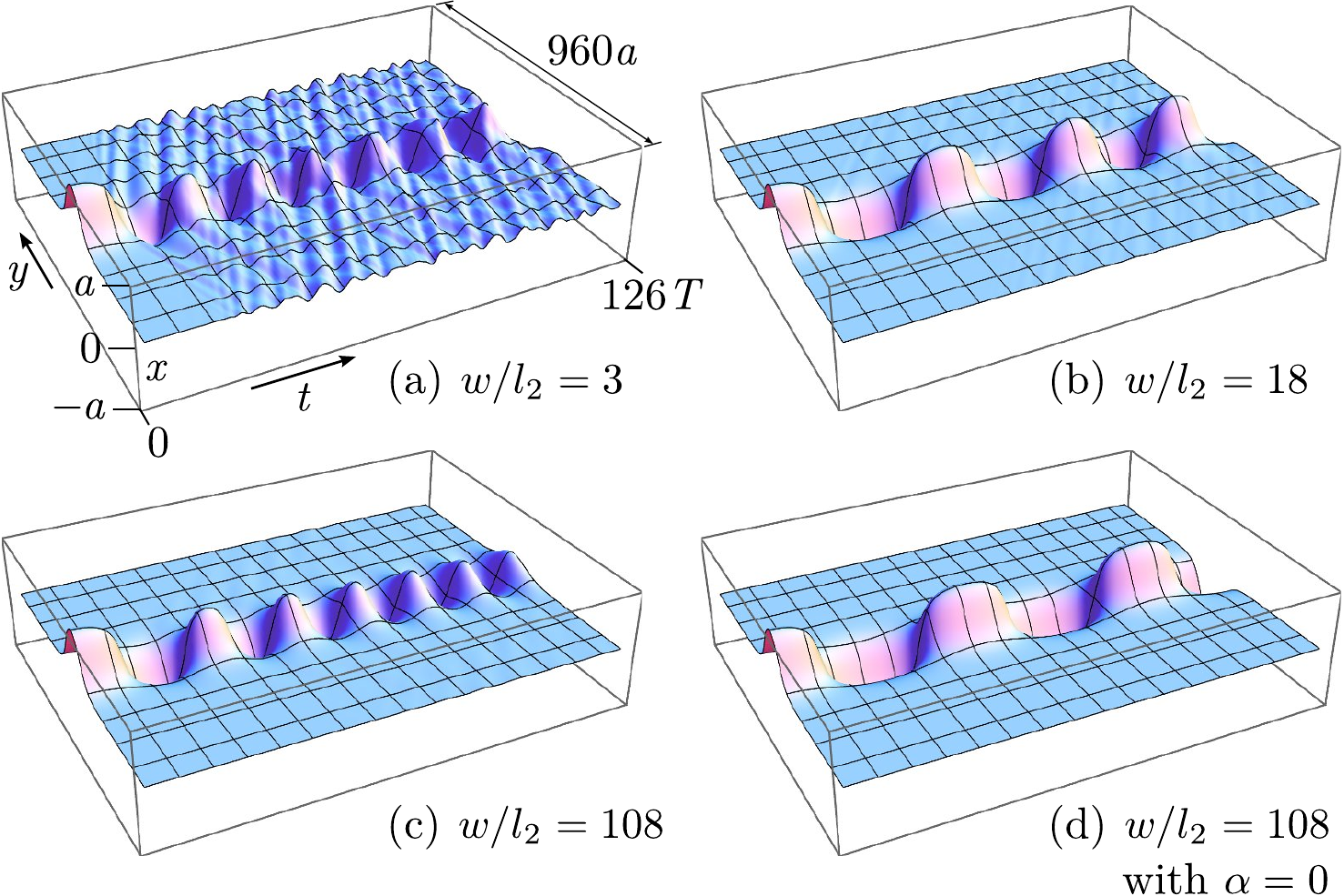}
  \caption{\label{fig:breathers}(color online). Domain-wall profiles $x(y,t)$ extracted from atomistic simulations of a breather ($w \approx 79a$, $\omega=0.25$).
  (a)~If $w \sim l_2$, the sine-Gordon picture of DW-kinks is inapplicable. The breather loses energy through spin-wave emission.
  (b,c)~Spin-wave emission is virtually absent for $w \gg l_2$. However, for high $w/l_2$ the breather is more susceptible to Gilbert damping ($\alpha = 0.0004$), resulting in a faster-decreasing amplitude and period.
  (d)~Solitonic limit ($w \gg l_2$ and no Gilbert damping). A video of the atomistic simulation is available \cite{supmat}.
  }
\end{figure}

First, $\vartheta$ must remain small at all times.
For $w \gg l_2$ and $\alpha = 0$, we have that $T\,\partial \varphi / \partial t \approx 2 l_1 w / (a l_2)\,\vartheta$.
A two-kink breather solution of \equaref{eq:sG}, similar to \figref{fig:breathers}(d), is given by \cite{Dodd1982}
\begin{equation}\label{eq:breathersol}
\varphi(y,t) = 4 \arctan \bigl[ \tfrac{\sqrt{1 - \omega^2}}{\omega} \sech \bigl( \tfrac{\sqrt{1 - \omega^2}y}{Y} \bigr) \cos \bigl( \tfrac{\omega t}{T} \bigr) \bigr] \text{,}
\end{equation}
where $\omega \in (0,1)$ is a parameter.
Notice that $| \partial \varphi / \partial t |$ attains a maximum at $y=0$, $t=\pi/(2\omega)\,T$, 
where the two kinks collide.
In the limit $\omega \rightarrow 0$, the breather \eqref{eq:breathersol} is equivalent to the collision of two nearly free kinks of opposite signs with negligible initial velocities.
We find $T|\partial \varphi / \partial t|_\text{max} = 4$ and $\vartheta_\text{max} = 2 a l_2 / (l_1 w)$.
Since typically $2a/l_1 \sim 1$, we conclude that the small-$\vartheta$ approximation is valid for
\begin{equation}\label{eq:condtheta}
w \gg l_2 \text{.}
\end{equation}
This condition must also be assumed to neglect the term in $\partial \vartheta / \partial y$ in \equaref{eq:hamapprox}.

Second, Gilbert damping must not be too strong.
We estimate the energy dissipated in a collision, treating Gilbert damping as a perturbation.
For $\alpha > 0$, \equaref{eq:sG} becomes
\begin{equation}\label{eq:sGdamped}
T^2 \frac{\partial^2 \varphi}{{\partial t}^2} - Y^2 \frac{\partial^2 \varphi}{{\partial y}^2} + \sin \varphi = - \xi T \frac{\partial \varphi}{\partial t} \text{,}
\end{equation}
where $\xi = \alpha w / (\pi l_2)$ is a dimensionless damping rate.
The energy dissipated in half a period of the breather (one collision) is given by
\begin{equation}
\Delta \mathcal{H} = -\xi \frac{E T}{Y} \int_{0}^{\pi T / \omega} \! \int_{-\infty}^{\infty}  {\Bigl(\frac{\partial \varphi}{\partial t}\Bigr)}^2\,dy\,dt \text{.}
\end{equation}
Substituting the original solution~\eqref{eq:breathersol}, which has an energy of $16 E \sqrt{1-\omega^2}$, we find that
$\Delta \mathcal{H} = -16 \xi E r(\omega) \sqrt{1-\omega^2}$, where $r(\omega)$ is a monotonic function with $r(0) = \pi^2/2$ and $r(1) = \pi$.
The relative energy loss for $\omega \rightarrow 0$ is thus given by $D = \pi \alpha w / (2l_2)$, and Gilbert damping may be considered small if
\begin{equation}\label{eq:condalpha}
\alpha w \ll l_2 \text{.}
\end{equation}
This condition is consistent with \equaref{eq:condtheta} only in materials with a very low Gilbert damping parameter $\alpha$.

\begin{figure}
  \includegraphics[width=\columnwidth]{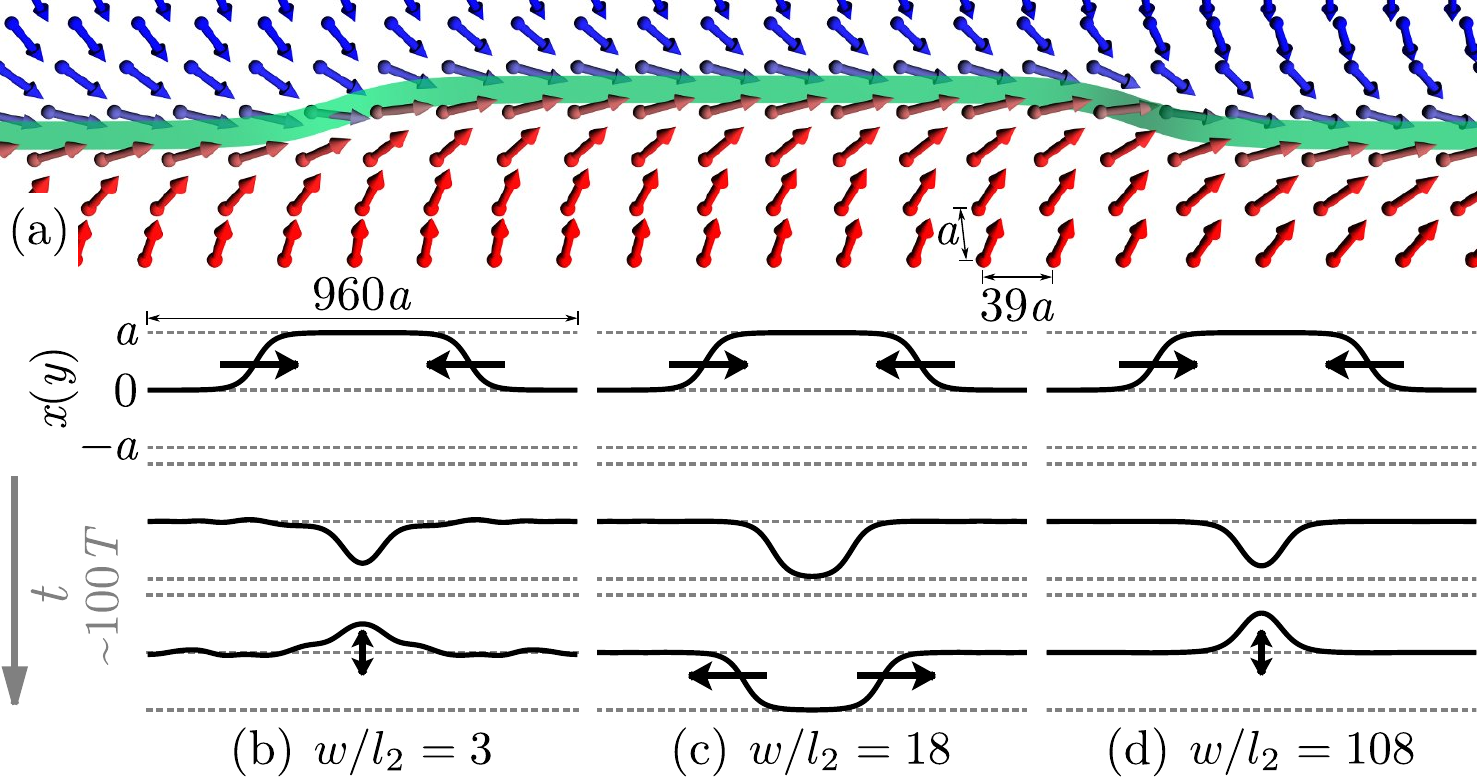}
  \caption{\label{fig:collisions}(color online).
  Simulations of kink collisions with initial velocities of $\pm 0.2 Y/T$ ($\alpha = 0.0004$, $l_1=1.58a$, $w \approx 79a$).
  (a)~Initial configuration. We extract the domain-wall profile $x(y)$ (strip) from the atomistic simulations.
  Not all magnetic moments are shown.
  (b)~For $w \lesssim l_2$, colliding kinks annihilate under emission of Winter spin waves \cite{Winter1961}.
  (c)~If conditions \eqref{eq:condtheta} and \eqref{eq:condalpha} are both satisfied, colliding kinks pass through each other. A segment of the domain wall makes a jump of distance $2a$ into another \Peierls{} valley, and propagation continues.
  (d)~For $\alpha w \gtrsim l_2$, colliding kinks lose energy through Gilbert damping. Like in (b), they become trapped in a breather and eventually annihilate.
  }
\end{figure}
For comparison, we perform atomistic spin-dynamics simulations, where we generate an initial configuration containing a domain wall with a two-kink profile, as shown in \figref{fig:collisions}(a), and numerically integrate the LLG equation~\eqref{eq:llg} for the Hamiltonian \eqref{eq:hamiltonian}.
We use the C++ code we developed with the implicit-midpoint integration scheme, verifying convergence of our results.
We extract the domain-wall profiles $x(y)$, shown in \figref{fig:collisions}(b-d), from the evolving atomistic spin configurations.
These results confirm that kinks may display solitonic behavior if the conditions~\eqref{eq:condtheta} and~\eqref{eq:condalpha} are satisfied.
\Figref{fig:breathers} shows that long-lived breathers can be observed under the same conditions.

For a crystal with uniaxial, perpendicular anisotropy ($K_2$ purely magnetostatic), we have
$l_2 = M_\text{S}^{-1} \sqrt{A / (2\pi)}$.
With parameter values from \explcite{Novoselov2003}, we get 
$l_2 = 0.11\,\text{{\textmu}m} \approx 61 a$ and $w / l_2 \approx 15$, so that \equaref{eq:condtheta} is satisfied. We remark that, while uniaxial anisotropy is dominant in thin films of bismuth- and gallium-substituted YIG \cite{Novoselov2003,Hansen1983}, there will be an additional contribution to $K_2$ from in-plane crystalline anisotropy. The extremely low Gilbert damping in pure YIG \cite{Serga2010} suggests that \equaref{eq:condalpha} may also be satisfied and that breathers could survive for many periods.

\paragraph*{Equations of motion.}

\begin{figure}
  \includegraphics[width=\columnwidth]{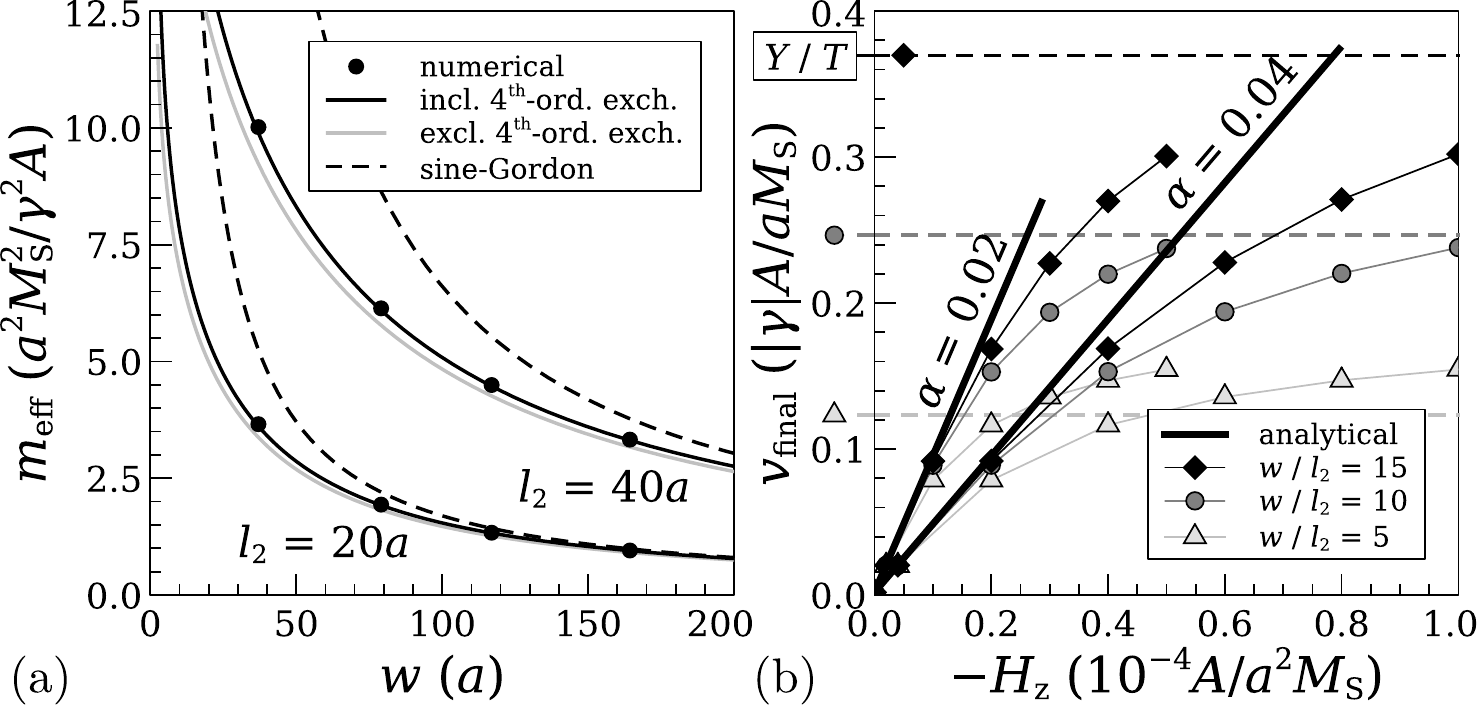}
  \caption{\label{fig:massvelo}
     (a) Our analytical expression \eqref{eq:effmass} for the kink effective mass $m_\text{eff}$ is in very good agreement with the numerical values for the atomistic model.
     Small corrections result from higher-order exchange terms \cite{supmat}.
     For $w \lesssim l_2$, $m_\text{eff}$ is reduced with respect to the sine-Gordon value $m_\text{sol}$.
     (b) Final velocity $v_\text{final}$ for $\alpha = 0.02$ and $\alpha = 0.04$ and for three values of $w/l_2$ ($w \approx 79a$).
     In the regime $v_\text{final} \ll Y/T$, $v_\text{final}$ follows \equaref{eq:vfinal} (solid lines) and is independent of $w/l_2$.
     Deviations occur when $v_\text{final}$ becomes comparable to $Y/T$ (horizontal lines).
     Unlike in the sine-Gordon model, the kink velocity can exceed $Y/T$.
  }
\end{figure}

A sine-Gordon soliton possesses inertia; its rest mass is given by $8 E T^2 / Y^2$ \cite{Dodd1982}. For DW-kinks, this evaluates to a mass of
\begin{equation}\label{eq:masssG}
m_\text{sol} = \frac{2 a^2 M_\text{S}^2}{\pi \gamma^2 K_2 l_1 w}
\end{equation}
per unit length.
We now derive nonrelativistic equations of motion valid for solitonic and nonsolitonic DW-kinks.
We linearize the Hamiltonian \eqref{eq:hamapprox} near a single kink at rest, for which we take $\vartheta(y)=0$ and $x(y)$ as in \equaref{eq:kinksolution} with $y_0=0$.
An inertial zero-frequency normal mode \cite{Buijnsters2014} is associated with the collective coordinate $y_0$.
We have $\partial x(y) / \partial y_0 = -(a / w) \sech ( \pi y / w )$ and $\partial \vartheta(y) / \partial y_0 = 0$.
We introduce a momentum $p$ and require that $p$ and $y_0$ decouple to second order from the other degrees of freedom. From $\dot{y}_0 = p/m_\text{eff}$, we derive $\partial x(y) / \partial p = 0$ and
\begin{equation}
m_\text{eff} \frac{\partial \vartheta(y)}{\partial p} = 
- \frac{a M_\text{S}}{2 |\gamma| K_2 l_1 w} {\Bigl(  1 - l_2^2 \frac{d^2}{{dy}^2} \Bigr)}^{-1} \sech \frac{\pi y}{w} \text{.}
\end{equation}
The effective mass $m_\text{eff}$ is fixed by the requirement that $y_0$ and $p$ be canonically conjugate, $\lbrace y_0, p \rbrace = 1$:
\begin{equation}\label{eq:effmass}
m_\text{eff} = f(w/l_2) \, m_\text{sol} \text{,}
\end{equation}
where we define
\begin{equation}\label{eq:feta}
f(\eta) = \frac{1}{2} \int_{-\infty}^\infty \frac{\sech^2 x}{1 + 4 x^2 / \eta^2} dx\text{.}
\end{equation}
For $\eta\gg 1$, $f(\eta) = 1 - \pi^2/(3\eta^{2}) + \Or(\eta^{-4})$.
In \figref{fig:massvelo}(a), we compare \equaref{eq:effmass} to the effective kink masses obtained for the atomistic model \eqref{eq:hamiltonian} using the numerical method described in \explcite{Buijnsters2014}. We find a very good agreement.

We introduce a characteristic angle $\vartheta_0$ related to the kink momentum $p$ via $\vartheta_0 = \pi |\gamma| / (4 a M_\text{S}) \, p$.
We derive from \equaref{eq:geneom} the linearized equations of motion for the collective coordinates $\vartheta_0$ and $y_0$,
\begin{subequations}\label{eq:kinkeom}
\begin{align}
\label{eq:kinkeomp}
\dot{\vartheta}_0& = -\frac{\pi}{2} |\gamma| H_z - \frac{\alpha}{R} \dot{y}_0 \text{,} \\
\label{eq:kinkeomy}
\dot{y}_0& = \frac{|\gamma| R}{f(\eta)} \Bigl[
  \Bigl(
  \frac{2  K_2 }{M_\text{S}}
  - \frac{\pi H_y }{2 g(\eta)}
  \Bigr) \vartheta_0
  + \frac{\pi^2}{4} H_x
  + \frac{\alpha}{|\gamma| g(\eta)} \dot{\vartheta}_0 \Bigr]
\text{,}
\end{align}
\end{subequations}
where $\mathbf{H}_\text{app} = H_x \hat{\mathbf{x}} + H_y \hat{\mathbf{y}} + H_z \hat{\mathbf{z}}$ is the applied field, 
$\eta = w/l_2$,
$R = l_1 w / a$, and
$g(\eta) = 2 f(\eta) / [\int_{-\infty}^\infty \! {(1 + 4 x^2 / \eta^2)}^{-2} \sech^2 x \, dx]$.
We have $\vartheta(y_0) = - h(\eta) \vartheta_0$, where $h(0) = 2/\pi$ and $h(\infty) = 1$.
The condition $\dot{\vartheta}_0 = 0$ results in a final velocity
\begin{equation}\label{eq:vfinal}
v_\text{final}
  = - \frac{\pi}{\alpha} \Bigl( \frac{l_1}{2 a} \Bigr)  |\gamma| H_z w 
\text{.}
\end{equation}
Our simulations, shown in \figref{fig:massvelo}(b), confirm this expression in the regime that $\vartheta_0 \ll 1$ and $v_\text{final} \ll Y / T $.

\paragraph*{Conclusion and outlook.}

We have derived explicit conditions for solitonic behavior of DW-kinks, in terms of Gilbert damping $\alpha$ and the lengths $w$ and $l_2$: $1 \ll w/l_2 \ll 1/\alpha$. For certain YIG films these conditions appear to be satisfied. In the solitonic regime, long-lived breathers can exist, as confirmed by our atomistic spin-dynamics simulations.
The sharp peak in the dynamical magnetic susceptibility observed in \explcite{Novoselov2003}, which survives for some time when the applied field is switched off, might be related to the existence of such breathers
\cite{Novoselovprivcommun},
although more experimental investigations are needed.
We have found expressions for the main dynamical characteristics of kinks, including effective rest mass, \equaref{eq:effmass}, and limiting velocity, \equaref{eq:vfinal}, which apply both in the solitonic regime and beyond.
By combining a number of Hall probes \cite{Novoselov2003}, one might be able to track the motion of individual DW-kinks.
Given the size of DW-kinks ($\sim 1\text{ \textmu m}$), it is conceivable that optomagnetical stimuli could be used to create kink pairs.
Such techniques would open the way to manipulation of magnetic domain walls with atomistic precision.

We thank Kostya Novoselov 
for stimulating discussions.
This work is part of the research programme of the Foundation for Fundamental Research on Matter (FOM), which is part of the Netherlands Organisation for Scientific Research (NWO).

\onecolumngrid
\appendix*

\newpage
\section{Supplemental Material}

\newlength{\figwidthhalf}
\setlength{\figwidthhalf}{246.0pt}

\newcommand{\secref}[1]{Sec.~\ref{#1}}
\newcommand{\Secref}[1]{Section~\ref{#1}}
\newcommand{\figrefs}[1]{Figs.~\ref{#1}}
\newcommand{\Figrefs}[1]{Figures~\ref{#1}}

\renewcommand\thefigure{A.\arabic{figure}}

\subsection{Kinks in domain walls versus kinks in dislocations}

We show that under certain conditions, DW-kinks may display solitonic behavior.
It is interesting to compare the dynamics of DW-kinks to kinks in dislocations in the crystal lattice.
\Explcite{Gornostyrev1999} found that solitonic behavior can be observed for dislocation kinks in a two-dimensional rigid-substrate model of the slip plane.
However, \Kosevich{} argues that the sine-Gordon equation cannot be a satisfactory model for the dynamics of a free dislocation because dislocation motion couples in an essential way to lattice vibrations \cite{Kosevich2005}.
Moreover, it seems that ballistic effects (let alone solitonic behavior) play no role of significance in practical simulations of kink or dislocation dynamics \cite{Swinburne2013A}.
It is important to note that a real dislocation is a line defect in three-dimensional space; its spatial profile depends in an essential way on the perpendicular coordinate. A domain wall, by contrast, is a planar defect. This justifies our approximation of a magnetic thin film containing a domain wall as effectively two-dimensional.

\subsection{Atomistic simulation of a breather (movie)}

\begin{figure}[h]
  \includegraphics[width=0.5\textwidth]{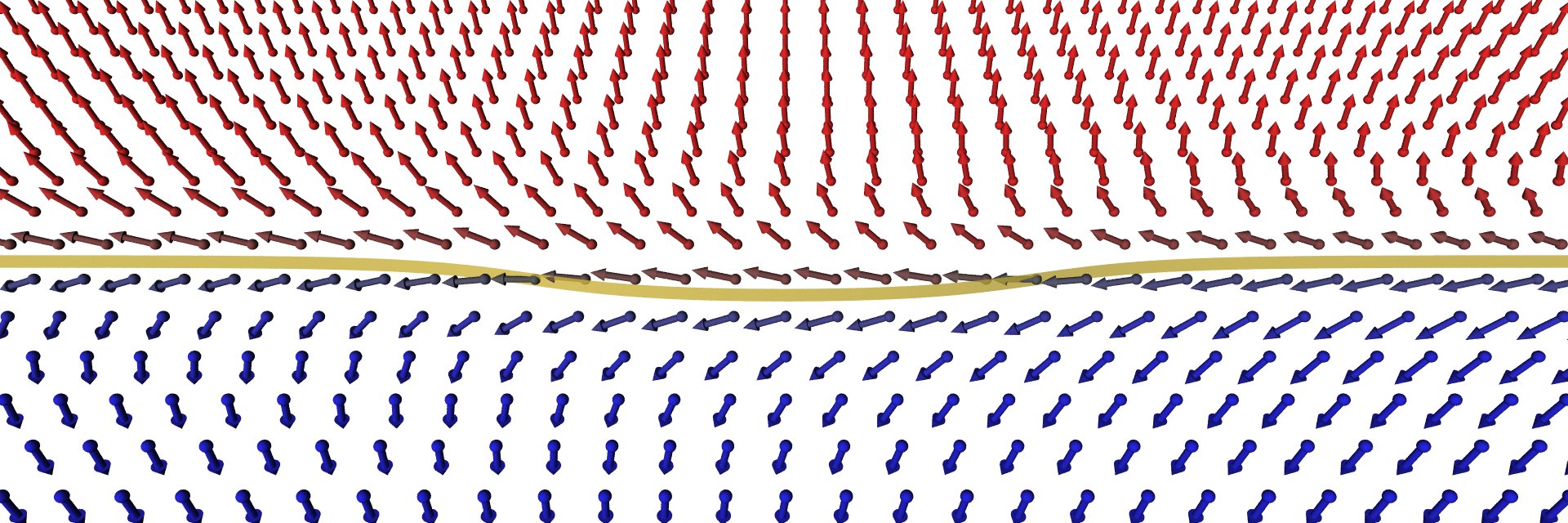}
  \caption{\label{fig:snapshot}(color online). Atomistic simulation (movie snapshot).}
\end{figure}

A movie file with an atomistic simulation of a breather is provided with this Supplemental Material (see \figref{fig:snapshot}). It shows a single period of a breather with $\omega = 0.1$ for $l_1 = l_2 = 1.58a$, $w \approx 79a$, and no Gilbert damping ($\alpha = 0$). The simulation box contains $40 \times 960$ magnetic moments.
The translucent yellow strip indicates the evolution of the domain-wall profile according to the analytical breather solution of the sine-Gordon equation.

\subsection{\label{sec:hoexch}Higher-order exchange}

\subsubsection{Continuum model}

For magnetization profiles smooth on the scale of the lattice constant $a$, the atomistic Hamiltonian (see main text) is equivalent to the sum of the continuum energy functionals
\begin{subequations}\label{eq:contenfunc}
\begin{align}
E_\text{ex}[\mathbf{m}(x,y)]& = \int \! A \biggl[ {\biggl\lVert \frac{\partial \mathbf{m}}{\partial x} \biggr\rVert}^2 + {\biggl\lVert \frac{\partial\mathbf{m}}{\partial y} \biggr\rVert}^2 \biggr] \, dx \, dy\text{,}\\
E_\text{ani}[\mathbf{m}(x,y)]& = \int  \! \left[ -K_1 {(\mathbf{m} \cdot \hat{\mathbf{z}})}^2 + K_2 {(\mathbf{m} \cdot \hat{\mathbf{x}} )}^2 \right] \, dx \, dy\text{,}\\
E_\text{Zee}[\mathbf{m}(x,y)]& = \int \! - M_\text{S} \mathbf{H}_\text{ext} \cdot \mathbf{m}\, dx \, dy\text{,}
\end{align}
\end{subequations}
where $\mathbf{m}$ represents as before the \emph{reduced} magnetization ($\lVert\mathbf{m}(x,y)\rVert = 1$).
If the characteristic length scale of the magnetization profile, set by the exchange length $l_1$, becomes comparable to $a$, the continuum energy functionals~\eqref{eq:contenfunc} are no longer a good representation of the atomistic Hamiltonian and correction terms are needed.
For our purposes, the most important correction is the \Peierls{} potential, which breaks translational symmetry. The \Peierls{} potential vanishes exponentially fast in $l_1 / a$.
However, there are also corrections to $E_\text{ex}$ that are \emph{algebraically} small. These can be expressed in terms of the higher-order spatial derivatives of $\mathbf{m}(x,y)$.
The dominant correction is
\begin{equation}\label{eq:fourthexch}
E_\text{ex4}[\mathbf{m}(x,y)] = \int \! B \biggl[ {\biggl\lVert \frac{\partial^2\mathbf{m}}{{\partial x}^2} \biggr\rVert}^2 + {\biggl\lVert \frac{\partial^2\mathbf{m}}{{\partial y}^2} \biggr\rVert}^2 \biggr] \, dx \, dy\text{.}
\end{equation}
While the usual exchange parameter $A$ is proportional to the second moment of the atomistic exchange kernel, the parameter $B$ is proportional to the fourth moment. For nearest-neighbor exchange in a square lattice, we derive
\begin{equation}\label{eq:secondexch}
B = - a^2 A / 12
\text{.}
\end{equation}
Energy terms such as \equaref{eq:fourthexch} do not affect the qualitative features of domain-wall kinks, but they do give certain corrections to kink parameters. Since $l_1 \sim a$, we may not assume that such corrections can be neglected.
Notice that \equaref{eq:secondexch} is specific to systems with only nearest-neighbor exchange; crystals with wider exchange kernels are likely to have a larger parameter $B/A$, so that the corrections derived here may be more significant.

\subsubsection{Derivation of the domain-wall Hamiltonian}\label{sec:dwham}

We now write the Hamiltonian in terms of the collective coordinates $x_0(y),\vartheta(y)$ of the domain wall. We aim to derive the effective Hamiltonian of the domain wall from the atomistic Hamiltonian in a systematic way, and we apply the approximation that $\vartheta$ is small only as a final step. In this section, we write $x_0$ for the position of the center of the domain wall to distinguish it from the spatial coordinate $x$.

Let us express the reduced magnetization $\mathbf{m}(x,y)$ in spherical coordinates,
\begin{equation}\label{eq:coordsys}
\mathbf{m} = \sin(\theta)\cos(\phi)\hat{\mathbf{x}} + \sin(\theta)\sin(\phi)\hat{\mathbf{y}} + \cos(\theta)\hat{\mathbf{z}}\text{.}
\end{equation}
Notice that we use $\vartheta$ to denote the collective coordinate of the domain wall, while we use $\theta$ for the polar angle of the magnetization $\mathbf{m}$ at a given point in space.
We assume that $\phi(x,y) = \vartheta(y) - \pi / 2$; the azimuthal angle $\phi$ is constant in $x$ and depends only on $\vartheta(y)$. As a result, a domain wall with $\vartheta \neq 0$ is equivalent to a \Bloch{} domain wall ($\vartheta = 0$) but with a modified anisotropy.
The effective anisotropy $K$ that the domain wall experiences is given by
\begin{equation}
K = K_1+K_2\sin^2 \vartheta \text{,}
\end{equation}
and we can define an effective `exchange length'
\begin{equation}
l(\vartheta) = \sqrt{\frac{A}{K_1+K_2\sin^2 \vartheta}} = \frac{l_1}{\sqrt{1+(K_2/K_1)\sin^2 \vartheta}}\text{.}
\end{equation}
We may now write the Hamiltonian of the domain wall, very generally, as
\begin{equation}\label{eq:effhamfull}
\mathcal{H}[x_0(y),\vartheta(y)] =  \int \! \frac{4A}{l(\vartheta)} \biggl[ s_0(x_0,l(\vartheta)) +  \frac{1}{2} s_x(x_0,l(\vartheta)) {\left( \frac{\partial x_0}{\partial y} \right)}^2 + \frac{l(\vartheta)^2}{2} s_\vartheta(x_0,l(\vartheta)) {\left( \frac{\partial \vartheta}{\partial y} \right)}^2 + \frac{\pi^2}{24} s_l(x_0,l(\vartheta)) {\left( \frac{\partial l(\vartheta)}{\partial y} \right)}^2 \biggr] \, dy \text{.}
\end{equation}
Here $s_0$, $s_x$, $s_\vartheta$, and $s_l$ are dimensionless functions of $x_0/a$ and $l/a$ that remain to be determined.
The characteristic length scale on which the domain-wall variables $x_0(y),\vartheta(y)$ vary in $y$ is $w$. Since we do not take into account higher-order derivatives in $y$, the Hamiltonian~\eqref{eq:effhamfull} is valid up to corrections of $\Or((w/a)^{-2})$. This is acceptable because $w \gg a$ while the dominant energy scale $\lambda$ is of $\Or((w/a)^{-1})$.
For all four functions $s_0$, $s_x$, $s_\vartheta$, and $s_l$, the dependence on $x_0$ vanishes as $\sim e^{-\pi^2 l/a} \sim \Or{(w/a)^{-2}}$. As the factors $(\partial \cdot / \partial y)^2$ are already of order $\Or{(w/a)^{-2}}$, we may neglect the $x_0$-dependence of $s_x$, $s_\vartheta$, and $s_l$.
For $s_0$, we write
\begin{equation}
s_0(x_0,l) = s_0(l) - v(l)\cos\frac{2 \pi x_0}{a}\text{,}
\end{equation}
which defines the \Peierls{} potential. Higher \Fourier{} components may be neglected (see main text).
In conclusion, we get
\begin{multline}\label{eq:effhamfullpeierls}
\mathcal{H}[x_0(y),\vartheta(y)] =  \int \! \frac{4A}{l(\vartheta)} \biggl[
    s_0(l(\vartheta))
  + \frac{1}{2} s_x(l(\vartheta)) {\left( \frac{\partial x_0}{\partial y} \right)}^2
  - v(l(\vartheta))\cos\frac{2 \pi x_0}{a}\\
  + \frac{l(\vartheta)^2}{2} s_\vartheta(l(\vartheta)) {\left( \frac{\partial \vartheta}{\partial y} \right)}^2
  + \frac{\pi^2}{24} s_l(l(\vartheta)) {\left( \frac{\partial l(\vartheta)}{\partial y} \right)}^2 \biggr] \, dy \text{.}
\end{multline}
If we expand to second order in $\vartheta$ and neglect any corrections of $\Or({(w/a)}^{-2})$, \equaref{eq:effhamfullpeierls} becomes
\begin{equation}\label{eq:hamfullapprox}
\mathcal{H} \approx \frac{4A}{l_1} \int \! \biggl[ s_0(l_1) + \frac{1}{2}\frac{K_2}{K_1}(s_0(l_1) - l_1 s'_0(l_1)) \vartheta^2 + \frac{1}{2}s_x(l_1) {\biggl( \frac{\partial x_0}{\partial y} \biggr)}^2
 - v(l_1) \cos\frac{2\pi x_0}{a} + \frac{l_1^2}{2}s_\vartheta(l_1) {\biggl( \frac{\partial \vartheta}{\partial y} \biggr)}^2 \biggr] \, dy
\text{.}
\end{equation}
\Equaref{eq:hamfullapprox} is equivalent to the domain-wall Hamiltonian given in the main text except for the correction factors $(s_0(l_1) - l_1 s'_0(l_1))$, $s_x(l_1)$, and $s_\vartheta(l_1)$.
By definition, $V_0 = \epsilon_1 v(l_1)$.

\subsubsection{Perturbative calculation of the functions $s_0(l)$, $s_x(l)$, $s_\vartheta(l)$, and $s_l(l)$}

Let us consider a planar domain wall with collective coordinates $x_0,\vartheta$ and effective exchange length $l$. As is well known, in spherical coordinates~\eqref{eq:coordsys}, its equilibrium magnetization profile $\mathbf{m}(x)$, taking into account only the continuum energy functionals~\eqref{eq:contenfunc}, is given by
\begin{subequations}
\begin{align}
\theta(x)& = 2\arctan\Bigl[\exp\Bigl( \frac{x-x_0}{l} \Bigr)\Bigr] \text{,}\\
\phi(x)& = \vartheta - \pi / 2 \text{.}
\end{align}
\end{subequations}
We derive the effect of the fourth-order exchange term~\eqref{eq:fourthexch} on the equilibrium magnetization profile in first-order perturbation theory.
In spherical coordinates~\eqref{eq:coordsys} and under the assumption that $\phi$ is constant, we have the identities ${\lVert \mathbf{m}' \rVert}^2 = {(\theta')}^2$ and ${\lVert \mathbf{m}'' \rVert}^2 = {(\theta')}^4 + {(\theta'')}^2$,
where the prime denotes the derivative in $x$. (For constant $\theta$, we have ${\lVert \mathbf{m}' \rVert}^2 = {(\phi')}^2 \sin^2 \theta$.)
For a planar profile, the continuum energy terms~\eqref{eq:contenfunc} become
\begin{subequations}
\begin{align}
E_\text{ex}& = \int \! A {(\theta')}^2 \, dx \text{,}\\
E_\text{ani}& = \int  \! K \sin^2 \theta \, dx\text{,}
\end{align}
\end{subequations}
while the fourth-order exchange term~\eqref{eq:fourthexch} becomes
\begin{equation}
E_\text{ex4} = \int \! B [ {(\theta')}^4 + {(\theta'')}^2 ] \, dx
\text{.}
\end{equation}
We find that, to first order in $B$, the equilibrium configuration is given by
\begin{equation}
\theta(x) = 2\arctan\Bigl[\exp\Bigl( \frac{x-x_0}{l} \Bigr)\Bigr] + \frac{B}{Al^2}\,\theta_1\Bigl(\frac{x-x_0}{l}\Bigr)\text{,}
\end{equation}
where
\begin{equation}
\theta_1(\zeta) = -3\tanh\zeta\sech\zeta + \tfrac{1}{2}\zeta\sech\zeta\text{.}
\end{equation}

We calculate the coefficients in \equaref{eq:hamfullapprox} from the (perturbed) magnetization profile of the planar domain wall.
Notice that the functions $s_0(l)$, $s_x(l)$, $s_\vartheta(l)$, and $s_l(l)$ are normalized in such a way that they approach $1$ in the continuum limit $B/(Al^2)\rightarrow0$.
We have
\begin{subequations}
\begin{align}
s_0& = \frac{l}{4A} \int \! (A {(\theta')}^2 + K\sin^2 \theta + B[{(\theta')}^4 + {(\theta'')}^2] ) \, dx \text{,}\\
s_x& = \frac{l}{2} \int \! {\left\lVert\frac{\partial\mathbf{m}}{\partial x_0}\right\rVert}^2 \, dx = \frac{l}{2} \int \! {(\theta')}^2 \, dx
\text{,}\\
s_\vartheta& = \frac{1}{2l} \int \! {\left\lVert\frac{\partial\mathbf{m}}{\partial \vartheta}\right\rVert}^2 \, dx = \frac{1}{2l} \int \! \sin^2\theta \, dx
\text{,}\\
s_l& = \frac{6l}{\pi^2} \int \! {\left\lVert\frac{\partial\mathbf{m}}{\partial l}\right\rVert}^2 \, dx = \frac{6}{\pi^2 l} \int \! {(x-x_0)}^2 {(\theta')}^2 \, dx
\text{,}
\end{align}
\end{subequations}
where we treat $l$ as a constant when taking the derivative in $\vartheta$.
We get
\begin{subequations}
\begin{align}
s_0(l)& = 1 + \tfrac{1}{2} (B/A)l^{-2} + \Or{({(l/a)}^{-4})} \text{,} \\
s_x(l)& = 1 - \tfrac{3}{2} (B/A)l^{-2} + \Or{({(l/a)}^{-4})} \text{,} \\
s_\vartheta(l)& = 1 + \tfrac{3}{2} (B/A)l^{-2} + \Or{({(l/a)}^{-4})} \text{,} \\
s_l(l)& = 1 - (24/\pi^2 + \tfrac{1}{2}) (B/A)l^{-2} + \Or{({(l/a)}^{-4})} \text{.}
\end{align}
\end{subequations}

\subsubsection{Corrections to kink parameters}

For convenience, let us define
\begin{subequations}
\begin{align}
S_0& = s_0(l_1) - l_1 s'_0(l_1) = 1 + \tfrac{3}{2} (B/A) l_1^{-2} + \Or(l_1/a)^{-4} \text{,}\\
S_x& = s_x(l_1) = 1 - \tfrac{3}{2} (B/A) l_1^{-2} + \Or(l_1/a)^{-4} \text{,}\\
S_\vartheta& = s_\vartheta(l_1) = 1 + \tfrac{3}{2} (B/A) l_1^{-2} + \Or(l_1/a)^{-4} \text{,}
\end{align}
\end{subequations}
which are the correction factors that appear in \eqref{eq:hamfullapprox}. \Equaref{eq:secondexch} gives $S_0 \approx 1 - \tfrac{1}{8} (l_1/a)^{-2}$, $S_x \approx 1 + \tfrac{1}{8} (l_1/a)^{-2}$, and $S_\vartheta \approx 1 - \tfrac{1}{8} (l_1/a)^{-2}$ for a square lattice with nearest-neighbor exchange.

As for the statical kink parameters, the Hamiltonian \eqref{eq:hamfullapprox} gives us a sine-Gordon kink solution (see main text) with
\begin{equation}
w = \frac{a}{2}\sqrt{\frac{S_x}{v(l_1)}} = S_x^{1/2} \, w_\text{cont} \text{,}
\end{equation}
where the subscript ${}_\text{cont}$ denotes the value in the continuum model with only second-order exchange, as used in the main text.
The kink energy is given by
\begin{equation}
\lambda = \frac{4a}{\pi} \epsilon_1 \sqrt{S_x v(l_1)} = S_x^{1/2} \, \lambda_\text{cont} \text{.}
\end{equation}

\begin{figure}
  \includegraphics[width=\figwidthhalf]{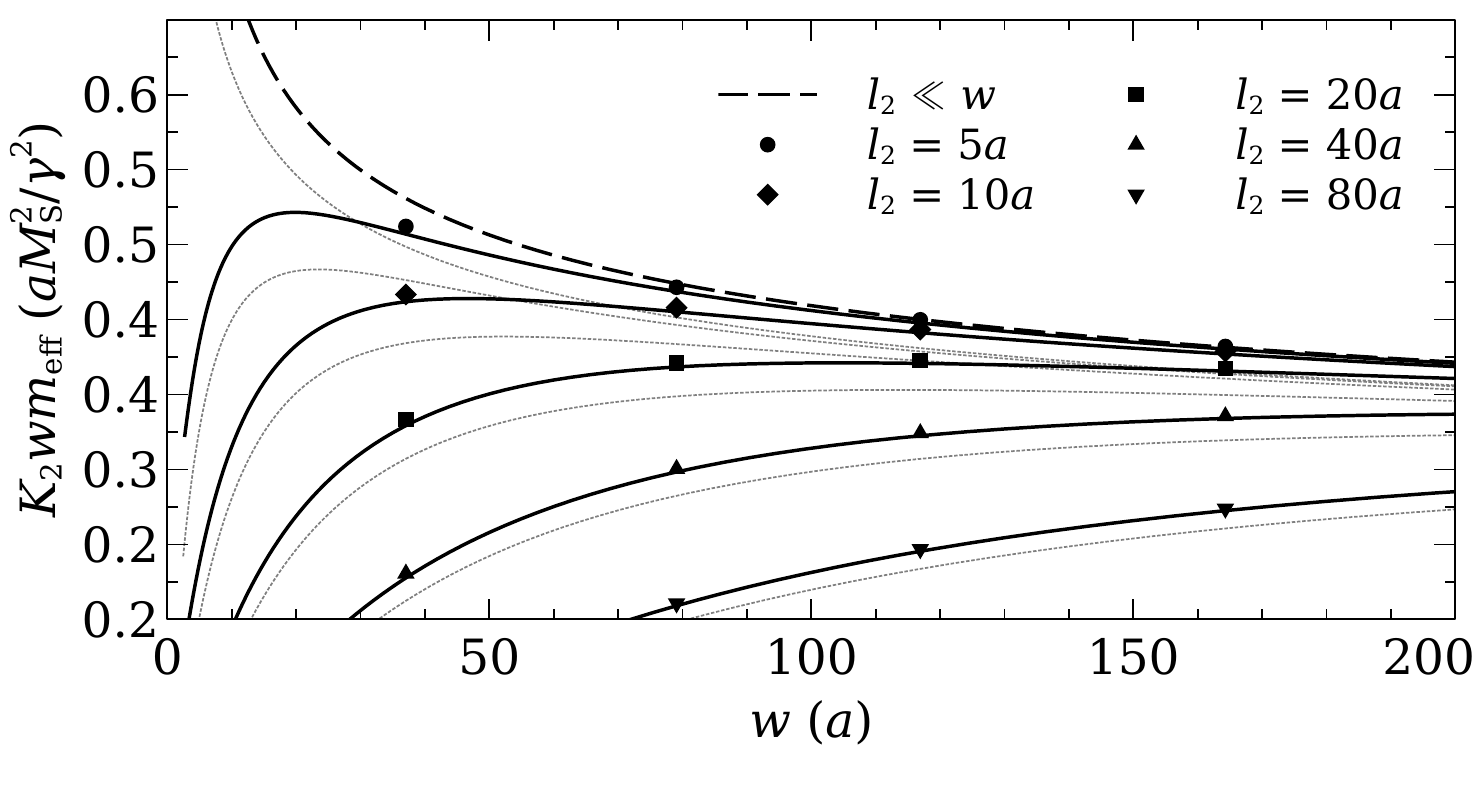}
  \caption{\label{fig:effmassfull}The analytical expression~\eqref{eq:meffcorr} for the effective mass $m_\text{eff}$ corrected for the effect of fourth-order exchange (solid black lines) is in almost perfect agreement with the numerical results for the atomistic model (symbols). The dotted lines show the uncorrected analytical expression given in the main text, which differs by $\sim 5\%$.
The dashed line shows the effective mass in the solitonic limit.
  We find that the (corrected) analytical expression deviates from the numerically calculated values only in the regime $w \lesssim 10a$ (not shown). This is not surprising because all our calculations assume $w \gg a$, as is always verified unless the \Peierls{} potential is extremely strong.}
\end{figure}
As for the dynamical kink parameters, we find
\begin{equation}\label{eq:meffcorr}
w m_\text{eff} = \frac{2 a^2 M_\text{S}^2 f(w/l_2)}{\pi \gamma^2 K_2 l_1 S_0} = \frac{f(w/l_2)}{f({(w/l_2)}_\text{cont})} S_0^{-1} \, {(w m_\text{eff})}_\text{cont}\text{,}
\end{equation}
where
\begin{equation}
l_2 = l_1 \sqrt{\frac{K_1}{K_2}} \sqrt{\frac{S_\vartheta}{S_0}} = S_0^{-1/2} S_\vartheta^{1/2} \, l_{2,\text{cont}} \text{.}
\end{equation}
\Figref{fig:effmassfull} shows that this correction to the effective mass $m_\text{eff}$ is small but significant. With the corrections, the agreement of our analytical expression to our numerical results is almost perfect.
The characteristic time scale of the approximate sine-Gordon description is given by
\begin{equation}
T = \frac{a M_\text{S}}{\pi |\gamma| \epsilon_1} \sqrt{\frac{K_1}{v(l_1) K_2 S_0}} = S_0^{-1/2} \, T_\text{cont} \text{.}
\end{equation}
The final velocity (in the linear regime, where $\vartheta_0 \ll 1$ and $v_\text{final} \ll Y / T$) is given by
\begin{equation}
\frac{v_\text{final}}{w} = S_x^{-1} \, {\Bigl(\frac{v_\text{final}}{w}\Bigr)}_\text{cont} \text{.}
\end{equation}

\subsection{Strength of the \Peierls{} potential}

The kink width $w$ depends on the strength $V_0$ of the \Peierls{} potential, which in turn depends on the ratio between the domain-wall width and the lattice periodicity $a$.
For the biaxial type of anisotropy defined in the main text, the equilibrium domain-wall width for a fixed value of $\vartheta$ is given by $\pi \sqrt{A / (K_1+K_2\sin^2 \vartheta)}$. For a \Bloch{} domain wall ($\vartheta=0$), this becomes $\pi l_1 = \pi \sqrt{A / K_1}$ \cite{Brown1963}.
Since we derive all analytical results in the limit of small $\vartheta$, $V_0$ can be taken as constant (see also \secref{sec:dwham}). It must be some dimensionless function of $l_1/a$ times the \Bloch{} domain-wall energy $\epsilon_1 = 4\sqrt{A K_1}$.

For a line of fixed $y = (j+\tfrac{1}{2})a$, we define the center $x$ of the domain wall as
\begin{equation}
x = \frac{a}{2} \sum_i (\mathbf{m}_{ij} \cdot \hat{\mathbf{z}})\text{,}
\end{equation}
which defines a position relative to the middle of the sample.
The magnetic \Peierls{} potential $V(x)$ can be obtained by minimizing the atomistic Hamiltonian of some configuration with a domain wall under the constraint of a fixed value of $x$.
We find, in agreement with previous considerations \cite{vandenBroek1971, Hilzinger1972},
\begin{equation}\label{eq:peierlspot}
V(x) = V_0  \left[ 1 - \cos \frac{2\pi x}{a} \right]\text{,}
\end{equation}
where $V$ is an energy per unit area.
We have $V_0 \sim \epsilon_1 e^{-\pi^2 l_1 / a}$ \cite{Hilzinger1972}.
For a noticeable \Peierls{} relief, we must have $l_1 \lesssim 2.5a$.
Higher \Fourier{} components of $V(x)$ decrease even faster in $l_1/a$ \cite{Kosevich2005}, so that \equaref{eq:peierlspot} is expected to describe the \Peierls{} potential almost perfectly unless $l_1 \lesssim 1.2a$. 
It is easy to see that an external field $\mathbf{H}_\text{app} = H_z \hat{\mathbf{z}}$ results in an additional potential $V_{H}(x) = -2 M_\text{S} H_z x$. Combining these two expressions gives a coercive field $H_\text{c} = \pi V_0 / (M_\text{S} a)$ \cite{Egami1973A}.
This provides an alternative way to determine numerically the strength $V_0$ of the \Peierls{} potential for a given atomistic model.

For a domain wall in a $\lbrace 1 0 0 \rbrace$ plane of the simple cubic lattice with nearest-neighbor exchange, we find that $V_0$ depends on the domain-wall width $l_1$ as
\begin{equation}\label{eq:peierlsstrength}
V_0 = (P\frac{l_1}{a} + Q) \epsilon_1 \, \e^{-\pi^2 l_1 / a}\text{.}
\end{equation}
A fit of the coercive fields gives $P \approx 181$ and $Q\approx-36$.
Our expression \eqref{eq:peierlsstrength} is equivalent to previous results \cite{Hilzinger1972,Novoselov2003} if we set $Q=0$ and $P=C/(4\pi)$.
We find that the refinement $Q\neq0$ is significant and makes the fit to the numerical results almost perfect.
If we consider a system with nearest-neighbor and next-nearest-neighbor exchange, \equaref{eq:peierlsstrength} remains valid but $P$ and $Q$ take different values. The \Peierls{} relief might thus be useful as a probe for the strength and type of the exchange interaction on the atomic scale.

\begin{figure}
  \centering
  \includegraphics[width=\figwidthhalf]{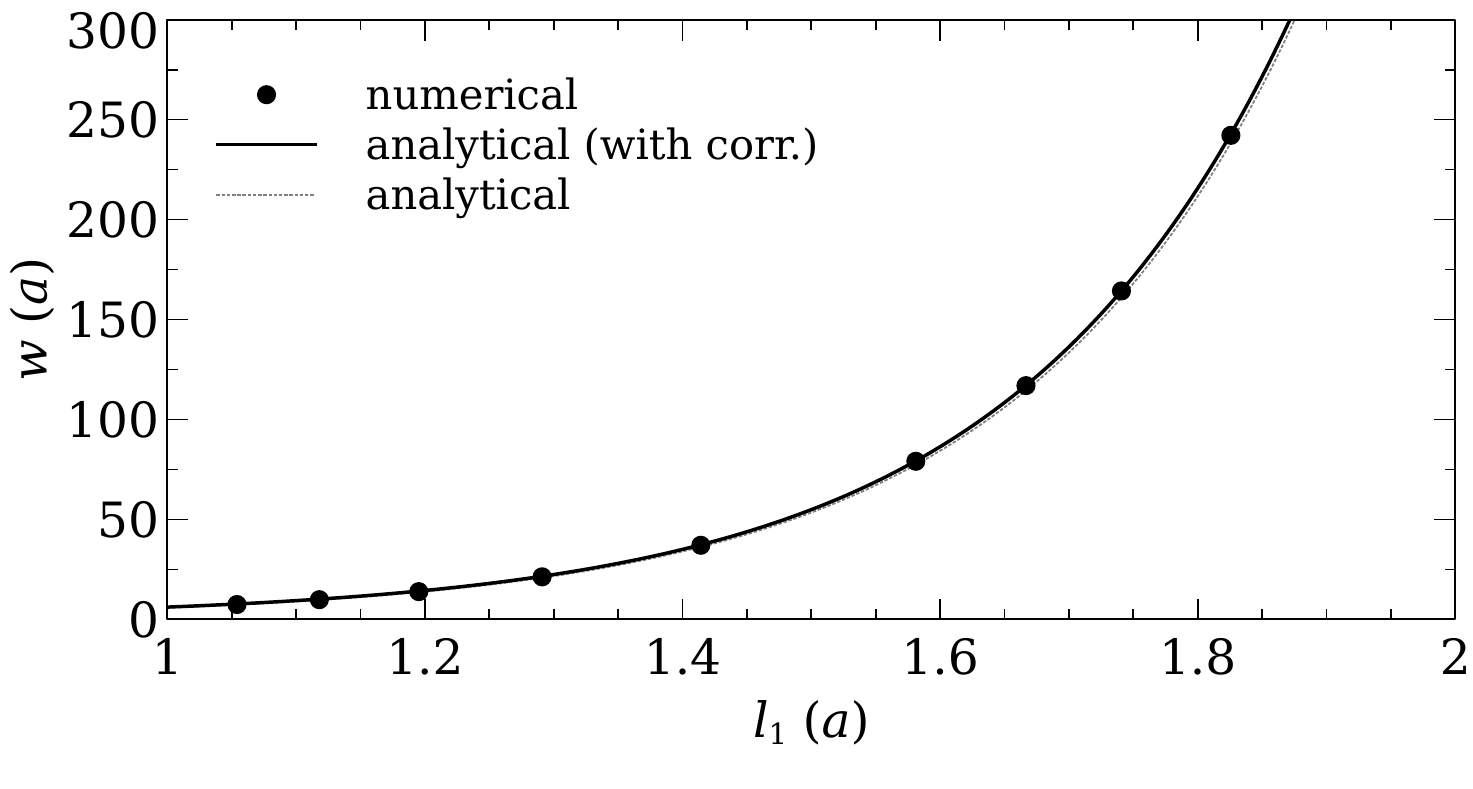}
  \caption{\label{fig:kinkwidth}Kink width $w$ versus exchange length $l_1$. Symbols:~Obtained by numerical minimization of the atomistic Hamiltonian (specified in the main text) starting from a spin configuration containing a domain wall with a single kink. The kink width is extracted from the domain-wall profile $x(y)$ in the final, relaxed configuration. Curves:~Calculated from \equaref{eq:peierlsstrength} using \equaref{eq:kinkwidth}, where the parameters $P$ and $Q$ are independently obtained from a fit to the coercive field.
  Notice that here the corrections to \equaref{eq:kinkwidth} that result from higher-order terms of the exchange energy (see \secref{sec:hoexch}) are tiny.
  }
\end{figure}
Given $V_0$, we can calculate the kink width $w$ using the relation \cite{Egami1973A}
\begin{equation}\label{eq:kinkwidth}
w = \tfrac{1}{2} a \sqrt{\epsilon_1 / V_0}\text{.}
\end{equation}
\Figref{fig:kinkwidth} shows that the kink widths calculated from \equaref{eq:peierlsstrength} agree with the results found by numerical relaxation of a domain wall with a kink.

\subsection{Adjoint forms of the kink collective coordinates}

In the main text, it is derived that the collective coordinates $y_0,p$ of the kink are given by
\begin{subequations}
\begin{align}
\frac{\partial x(y) }{ \partial y_0}& = -\frac{a}{w} \sech \frac{\pi y}{ w } \text{,}\\
\frac{\partial \vartheta(y) }{ \partial y_0}& = 0, \\
\frac{\partial x(y) }{ \partial p}& = 0\text{,}\\
\frac{\partial \vartheta(y)}{\partial p}& = 
- \frac{a M_\text{S}}{2 |\gamma| K_2 l_1 m_\text{eff} w} {\Bigl(  1 - l_2^2 \frac{d^2}{{dy}^2} \Bigr)}^{-1} \sech \frac{\pi y}{w} \text{.}
\end{align}
\end{subequations}
For completeness, we mention that, owing to the symplectic structure \cite{Buijnsters2014} on the domain-wall coordinates $x(y),\vartheta(y)$, these expressions immediately imply a `direct' definition of $y_0$ and $p$, namely
\begin{subequations}\label{eq:direct}
\begin{align}
y_0& = - \frac{\pi}{2a f(w/l_2)}  \int \! \Delta x(y) {\Bigl(  1 - l_2^2 \frac{d^2}{{dy}^2} \Bigr)}^{-1} \sech \frac{\pi y}{w} \, dy \text{,}\\
\label{eq:pdirect}
p& = - \frac{2 a M_\text{S}}{|\gamma| w} \int \! \vartheta(y)  \sech \frac{\pi y}{w} \, dy
 \text{.}
\end{align}
\end{subequations}
This definition is valid up to first order in the deviations $\Delta x(y), \vartheta(y)$ from the equilibrium kink configuration (for which $\vartheta(y)=0$ identically).
We have $\Delta x(y) = x(y) - (2a/\pi) \arctan [ \exp  (\pi y / w)  ]$.
A direct definition of collective coordinates is useful when deriving their equations of motion under some non-Hamiltonian perturbation such as Gilbert damping.
Notice that \equaref{eq:pdirect} gives
\begin{equation}
\vartheta_0 = \frac{\pi |\gamma|}{4 a M_\text{S}} \, p = - \frac{\pi}{2w} \int \! \vartheta(y)  \sech \frac{\pi y}{w} \, dy \text{.}
\end{equation}

\subsection{Special functions}

\begin{figure}
  \centering
  \subfloat{
    \includegraphics[height=0.5\figwidthhalf]{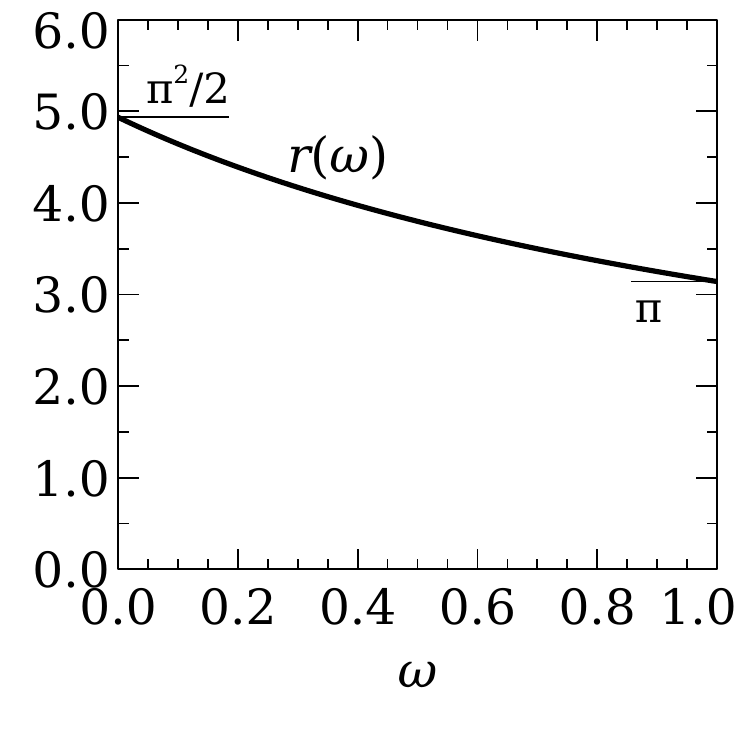}
  }

  \caption{\label{fig:romega}Special function $r(\omega)$, with limiting values.}
\end{figure}

\begin{figure}
  \subfloat{
    \includegraphics[height=0.5\figwidthhalf]{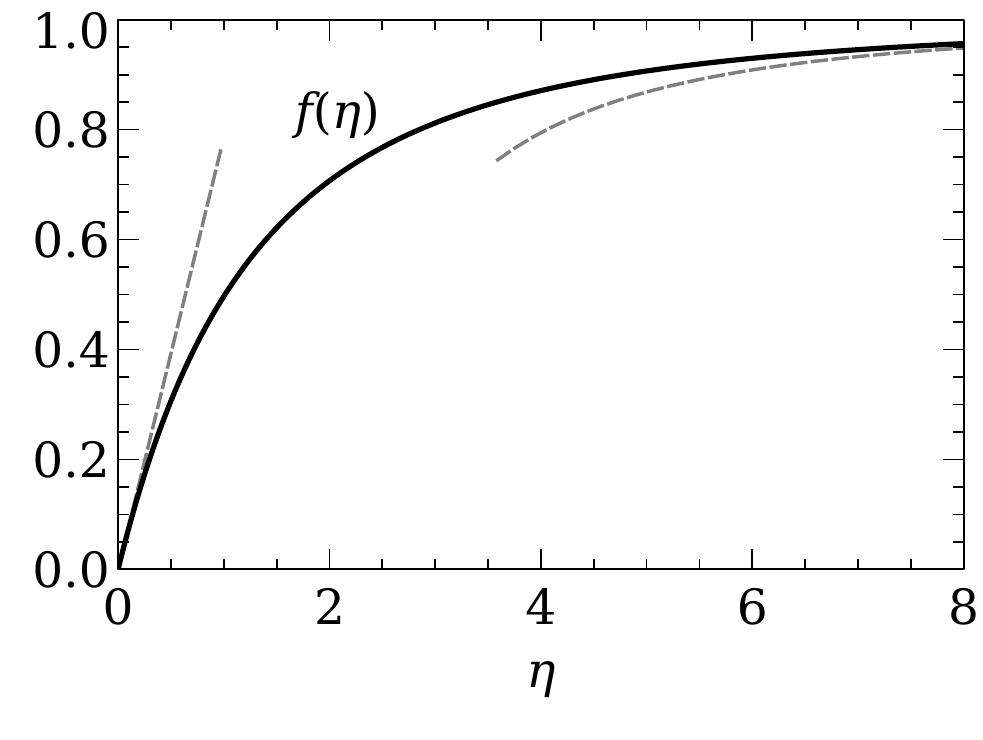}
  }
  \subfloat{
    \includegraphics[height=0.5\figwidthhalf]{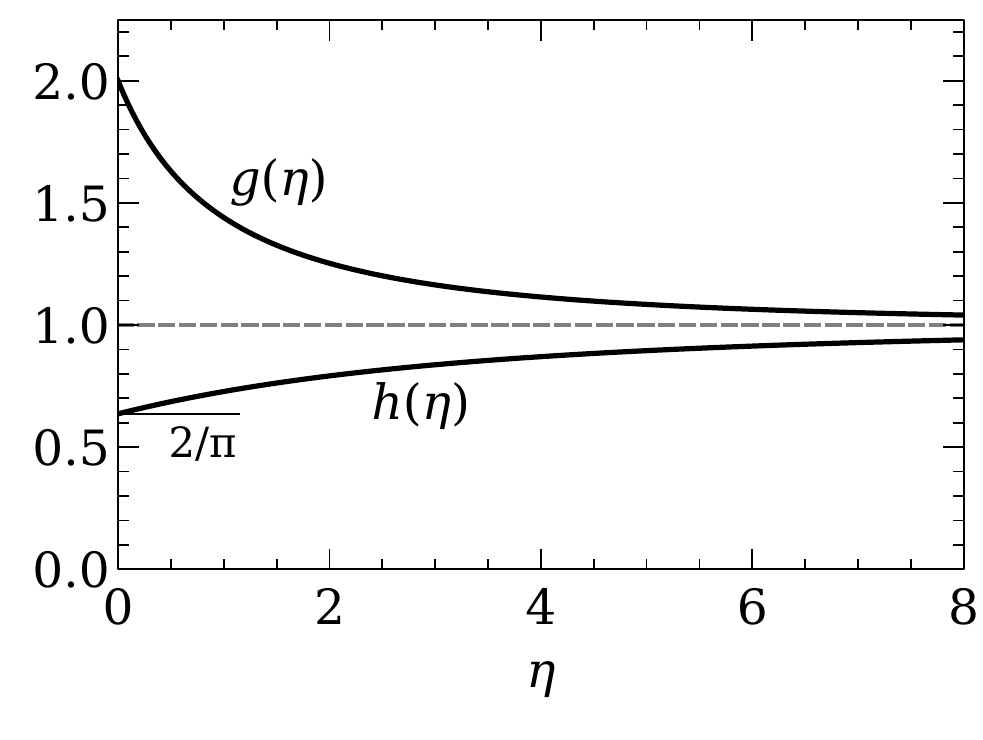}
  }

  \caption{\label{fig:fgheta}Special functions $f(\eta)$, $g(\eta)$, and $h(\eta)$, with limiting behavior.}
\end{figure}

The main text introduces the special functions $r(\omega)$, $f(\eta)$, $g(\eta)$, and $h(\eta)$. For completeness, we summarize their definitions and limiting behavior here. The functions are plotted in \figrefs{fig:romega} and~\ref{fig:fgheta}.
We have
\begin{equation}\label{eq:defr}
r(\omega) = \frac{1}{16\sqrt{1-\omega^2}} \int_{0}^{\pi/\omega} \! \int_{-\infty}^\infty \! {\left( \frac{\partial \varphi}{\partial t} \right)}^2 \, dy \, dt \text{,}
\end{equation}
where
\begin{equation}\label{eq:breathersol2}
\varphi(y,t) = 4 \arctan \left[ \frac{\sqrt{1 - \omega^2}}{\omega} \sech{\left( \sqrt{1 - \omega^2}y \right)} \cos{\left( \omega t \right)} \right]
\end{equation}
is a breather solution of the sine-Gordon equation, with parameter $\omega \in (0,1)$. We have made all variables dimensionless. The solution~\eqref{eq:breathersol2} is periodic in $t$ with a period of $2\pi / \omega$.
\Equaref{eq:defr} can be evaluated as
\begin{equation}\label{eq:evalr}
r(\omega) = \frac{2\pi}{\sqrt{1-\omega^2}} \arctan{\sqrt{\frac{1-\omega}{1+\omega}}} \text{.}
\end{equation}
It is easy to see that $r(\omega)$ is a bounded function with
\begin{subequations}
\begin{align}
r(\omega \downarrow 0)& = \pi^2 / 2 \approx 4.935 \text{,}\\
r(\omega \uparrow 1)& = \pi \approx 3.142 \text{.}
\end{align}
\end{subequations}
For $\eta > 0$ we define
\begin{equation}
f(\eta) = \frac{1}{2} \int_{-\infty}^\infty \! \frac{\sech^2{x}}{1 + 4 x^2 / \eta^2} \, dx \text{,}
\end{equation}
which has the following limiting behavior:
\begin{subequations}
\begin{align}
f(\eta)& = \tfrac{1}{4} \pi |\eta| + \Or(\eta^{2}) \enspace\text{for $\eta \ll 1$}\text{,}\\
f(\eta)& = 1 - \tfrac{1}{3}\pi^2/\eta^{2} + \Or(\eta^{-4}) \enspace\text{for $\eta \gg 1$}\text{.}
\end{align}
\end{subequations}
We also define two related functions
\begin{align}
g(\eta)& = 2 f(\eta) {\left[\int_{-\infty}^\infty \! \frac{\sech^2{x}}{{(1 + 4 x^2 / \eta^2)}^{2}}   \, dx\right]}^{-1}\text{,} \\
h(\eta)& = \frac{1}{\pi f(\eta)} \int_{-\infty}^\infty \! \frac{\sech{x}}{1+4x^2/\eta^2} \, dx \text{,}
\end{align}
for which we have
\begin{subequations}
\begin{align}
g(\eta \downarrow 0)& = 2 \text{,}\\
g(\eta)& = 1 + \tfrac{1}{3}\pi^2/\eta^{2} + \Or(\eta^{-4}) \enspace\text{for $\eta \gg 
1$}\text{,}
\end{align}
\end{subequations}
\begin{subequations}
\begin{align}
h(\eta \downarrow 0)& = 2/\pi \approx 0.6366 \text{,}\\
h(\eta)& = 1 - \tfrac{2}{3}\pi^2/\eta^{2} + \Or(\eta^{-4}) \enspace\text{for $\eta \gg 1$}
\text{.}
\end{align}
\end{subequations}
Notice that $f(\eta)$, $g(\eta)$, and $h(\eta)$ are defined in such a way that they approach~$1$ for large $\eta$ (solitonic limit).

\end{document}